\documentclass[aps,preprint,nofootinbib]{revtex4}
\usepackage{amsmath}
\usepackage{amssymb}
\begin{document}
\author{R. N. Ghalati}
\email{rnowbakh@uwo.ca}
\affiliation{Department of Applied Mathematics,
University of Western Ontario, London, N6A~5B7 Canada}
\title{\begin{flushright}
 {\footnotesize UWO\,-TH-\,07/7}
\end{flushright}Constraint Analysis of Linearized Gravity and a Generalization of the HTZ Approach}
\date{\today}
\begin{abstract}
The Dirac constraint formalism is applied to linearized gravity to determine the structure
of constraints and construct the canonical Hamiltonian. The diffeomorphism invariance of the Lagrangian is
retrieved by a nontrivial generalization of the method of Henneaux, Teitelboim and Zanelli, which takes into account 
the appearance of spatial derivatives of constraints in the constraint structure. 
A couple of first order formulations of the theory are discussed with the hope of opening avenues on an 
unambiguous canonical treatment of the Einstein-Hilbert action in its first order form.   
\end{abstract}
\maketitle
\section{Linearized Gravity and the Constraint Structure}
The metric $g_{\mu\nu}$ of the gravitational field in an $n+1$ dimensional 
spacetime\footnote{The ``mostly plus \cite{cliff}'' metric is used. Greek and Latin indices
stand for spacetime and spatial coordinates respectively.} 
is known to extremize the Einstein-Hilbert action as a functional of the metric and its first and second order derivatives,
\begin{equation}
\label{001}
S_{EH}=\int d^{n+1}x\,\, \mathcal{L}\,.
\end{equation}
Here we have the Lagrangian density
\begin{equation}
\label{002}
\mathcal{L}
=\sqrt{-g}\,g^{\mu\nu}\left(\Gamma^\lambda_{\mu\nu,\lambda}-\Gamma^\lambda_{\lambda\mu,\nu}+
\Gamma^\lambda_{\mu\nu}\Gamma^\sigma_{\sigma\lambda}-\Gamma^\lambda_{\sigma\mu}\Gamma^\sigma_{\lambda\nu}\right)
\end{equation}
where
\begin{equation}
\label{003}
\Gamma^\lambda_{\mu\nu}=\frac{1}{2}\,g^{\lambda\sigma}\left(g_{\sigma\mu,\nu}+g_{\sigma\nu,\mu}-g_{\mu\nu,\sigma}\right)
\end{equation}
are the Christoffel symbols and $g=\det g_{\mu\nu}$. The action for the `` weak field limit'' of the 
gravitional field equations is defined as the Einstein-Hilbert action for {\it small} 
perturbations around a fixed background potential, neglecting terms of higher order in perturbations. 
The latter implies that the perturbations are tensors under the isometry group of the fixed background 
spacetime. This means for instance that raising and lowering of indices for the perturbation field are
 done by the background metric field.

If the Minkowski metric $\eta_{\mu\nu}$ is the background, then the weak field limit of the Einstein-Hilbert 
Lagrangian is achieved in the following way.
If  $\phi_{\mu\nu}$ are perturbations around $\eta_{\mu\nu}$,
\begin{equation}
\label{004}
g_{\mu\nu}=\eta_{\mu\nu}+\phi_{\mu\nu}\,,
\end{equation}
then
\begin{eqnarray}
\label{005}
g^{\,\mu\nu}&=&\eta^{\,\mu\nu}-\phi^{\,\mu\nu}+O\,(\phi^2)\\
\Gamma^\lambda_{\mu\nu}&=&\frac{1}{2}\left(\,\phi^{\,\,\,\lambda}_{\mu\,\,\,,\,\nu}+
\phi^{\,\,\,\lambda}_{\nu\,\,\,,\,\mu}-\phi^{\,\,\,\,\,\,\,,\,\lambda}_{\mu\nu}\,\right)+O\,(\phi^2)\,\,,
\end{eqnarray}
and keeping only the contributions to eqn. (\ref{002}) that are bilinear in $\phi_{\mu\nu}$ 
the Einstein-Hilbert Lagrangian density goes over into 
\begin{equation}
\label{006}
\mathcal{L}=\frac{1}{4}\,\partial_\mu {\phi^\nu}_\nu \,\partial^\mu {\phi^\lambda}_\lambda-
\frac{1}{4}\,\partial_\lambda \phi_{\mu \nu}\,\partial^\lambda \phi^{\mu \nu}+
\frac{1}{2}\,\partial_\mu{\phi^\mu}_\nu \,\partial_\lambda \phi^{\lambda \nu}-
\frac{1}{2}\,\partial_\mu \phi^{\mu \nu}\,\partial_\nu {\phi^\lambda}_\lambda\,.
\end{equation}
This action, which is the action for a spin-2 field $\phi_{\mu\nu}$ in an $n+1$ dimensional flat spacetime, 
can also be derived by alternative ways such as by use of the Bargmann-Wigner equation \cite{Gerry}.

Assuming lower index fields $\phi=\phi_{00}$, $\phi_i=\phi_{0i}$ and $\phi_{ij}=\phi_{ij}$ to be dynamical, the canonical momenta 
\begin{equation}
\label{113}
\pi=\frac{\delta L}{\delta \dot \phi}\,\,\,\,\,\,\,\,\,\,\,\,\,\,\,\pi_i=\frac{\delta L}
{\delta \dot \phi_i} \,\,\,\,\,\,\,\,\,\,\,\,\,\,\,\pi_{ij}=\frac{\delta L}{\delta \dot \phi_{ij}}
\end{equation}
are  
\begin{eqnarray}
\label{113-1}
\pi &=& \frac{1}{2}\,\partial_k \phi_k\\
\pi_k &=& \frac{1}{2}\, \partial_k \phi_{ll}-\frac{1}{2}\,\partial_k \phi- \partial_l \phi_{lk}\,\,\,\,\,\,\,\,\,\,\,\,\,\,\,\,\,\, k=1\ldots n\\
\pi_{ij} &=& \frac{1}{2}\,[\,\dot \phi_{ij}+\delta_{ij}(-\dot \phi_{mm}+\partial_k \phi_k)\,]\,\,\,\,\,\,\,\,\,\,\,\,\,\,\,\,\,\, i,j=1\ldots n\,.
\end{eqnarray}
The last equation can be solved for $\dot \phi_{ij}$ in terms of $\pi_{ij}$ and the Hamiltonian density 
\begin{equation}
\label{1131-1}
\mathcal{H}=\pi \dot \phi +\pi_{k} \dot \phi_{k} + \pi_{ij} \dot \phi_{ij}-\mathcal{L}
\end{equation} 
is  
\begin{multline}
\label{1131}
\mathcal{H}=\pi_{ij}\pi_{ij}-\frac{1}{n-1}\pi_{mm}\pi_{nn}+\frac{1}{n-1}\pi_{mm}\partial_k\phi_k+\frac{n-2}{4(n-1)}\partial_k\phi_k \,\partial_l \phi_l \\+\frac{1}{2}\partial_m \phi \,\partial_m \phi_{ll}-\frac{1}{4}\partial_m\phi_{ll}\partial_m\phi_{kk}-\frac{1}{2}\partial_m \phi_k \partial_m \phi_k+\frac{1}{4}\partial_m\phi_{kl}\partial_m\phi_{kl}\\-\frac{1}{2}\partial_m\phi_{ml}\partial_n\phi_{nl}-\frac{1}{2}\partial_n \phi\, \partial_m\phi_{mn}+\frac{1}{2}\partial_m\phi_{mn}\,\partial_n\phi_{ll}\,,
\end{multline}
while by eqns\ (9) and (10) we have the $n+1$ primary $1^{st}$class constraints 
\begin{eqnarray}
\label{114} 
\Pi &=& \pi-\frac{1}{2}\,\partial_k\phi_k \approx 0\\
\Pi_k &=& \pi_k +\partial_l \phi_{lk}+\frac{1}{2} \,\partial_k\phi-\frac{1}{2}\,\partial_k\phi_{ll} 
\approx 0\,\,\,\,\,\,\,\,\,\,\,\,\,\,\,\,\,\,\,\,k=1\ldots n\,. 
\end{eqnarray}
The time change of these primary $1^{st}$class constraints 
\begin{equation}
\label{115-1}
\chi =\left\{\Pi,H\right\}\,\,\,\,\,\,\,\,\,\,\,\,\,\,\,\,\,\,\,\,\,\,\,\,\,\,\,\,\,\,\,\,\,\,\,\,\,\,\chi_k =\left\{\Pi_k,H\right\}, 
\end{equation}
leads to a set of $n+1$ secondary $1^{st}$class constraints 
 \begin{eqnarray}
\label{115} 
\chi &=& \frac{1}{2}\left(\partial_{mm}\phi_{ll}-\partial_{mn}\phi_{mn}\right)\approx 0\\
\chi_k &=& -\,\partial_{mm}\phi_k+2\,\partial_l\pi_{lk} \approx 0\,\,\,\,\,\,\,\,\,\,\,\,\,\,\,\,\,\,\,\,\ k=1\ldots n \,.
\end{eqnarray}  

All the $2(n+1)$ $1^{st}$class constraints are seen to commute. No tertiary constraint arises but 
interestingly enough spatial derivatives of secondary constraints appear when computing 
the time change of secondary constraints as 
\begin{equation}
\label{116} 
\left\{\chi,H\right\}=-\frac{1}{2}\,\partial_k \chi_k
\end{equation}
while
\begin{equation}
\label{117} 
\left\{\chi_k,H\right\}=0\,.
\end{equation} 
Since there are $(n+1)(n+2)/2$ independent fields $\phi_{\mu\nu}$ in $n+1$ dimensions and there are 
altogether $2(n+1)$ $1^{st}$class constraints on the system, then there are $(n+1)(n+2)/2-2(n+1)=\frac{1}{2}(n-2)(n+1)$ 
degrees of freedom in consfiguration space; zero degrees of freedom in three dimensional 
spacetime and two degrees of freedom in four dimensions.\footnote{In two dimensions when $n=1$, 
the Lagangian for linearized gravity reduces to a surface term and hence is not dynamical.}
 
To obtain the gauge transformations by means of $1^{st}$class constraints, an 
extension of the approach of Henneaux, Teitelboim and Zanelli \cite{Henneaux} is necessary as the appearance 
of spatial derivatives of constraints in the constraint structure such as in eqn. (\ref{116}) does not 
occur in their analysis of gauge transformations.    
\section{Extension of the HTZ approach}
We shall use notation of ref. \cite{Henneaux2} to make it easier for the reader already 
familiar with the approach to follow. Consider a system whose 
equations of motion are derived from an action
\begin{equation}
\label{1}
S=\int dt\, L\,,
\end{equation}
where $L$ is a local Lagrangian of {\it fields} $q^n=q^n\left(t,x^1,...,x^m\right)$ and their first order derivatives
\begin{equation}
\label{2}
L=\int d^mx \,\,\mathcal{L}\, \left(q^n,\partial_\mu q^n\right)\,,
\end{equation}
and assume that the Dirac constraint formalism has yielded the following first order Lagrangian
\begin{equation}
\label{3}
L=L\left(q^n,p_n\right)=\int d^mx \left(\,p_n{\dot q}^n-\mathcal{H}\left(q^n,p_n\right)\right)
\end{equation}
with the following constraint structure
\begin{eqnarray}
\label{4}
\left\{\gamma_a,\gamma_b\right\} &=& C_{ab}^c \gamma_c\\
\left\{H,\gamma_a\right\} &=& V_a^b \gamma_b+U_a^{cb} \partial_c \gamma_b\,,
\end{eqnarray}
where $p_n$ is the canonical momentum conjugate to $q^n$, $\gamma$'s represent $1^{st}$class constraints of all generations and 
$H=\int d^mx\, \mathcal{H}\left(q^n,p_n\right)$ is the canonical Hamiltonian of the system. The new feature is the 
appearance of the second term on the right hand side of eqn.\ (25). Such a term arises in eqn. (\ref{116}). 
Though we are not taking $2^{nd}$class constraints
into consideration, the analysis is general enough to include linearized gravity and most other 
Lagrangians that are physically relevant.

The extended action for the system then is 
\begin{equation}
\label{5}
S_E=S_E\left(q^n,p_n,u^a\right)=\int\! dt\, L_E
\end{equation} 
where 
\begin{equation}
\label{6}
L_E=\int d^mx\,\left(p_n{\dot q}^n-\mathcal{H}-u^a\gamma_a\right),
\end{equation}
incorporating all $1^{st}$class constraints of all generations by means of Lagrange multipliers $u^a$.
This makes the extended action a functional of not only $p_n$ and $q^n$ but also $u^a$. A gauge transformation on
 $F=F\left(q^n,p_n\right)$ is
\begin{equation}
\label{7}
\delta_\epsilon F= \epsilon^a\left\{F,\gamma_a\right\}\,\,,
\end{equation}
where $\epsilon^a=\epsilon^a\left(x^\mu\right)$ are arbitrary functions with proper behavior on the boundaries.
The extended action of eqn. (\ref{5}) is invariant up to a surface term 
under the gauge transformation of eqn. (\ref{7}) provided one simultaneousely transforms Lagrange multipliers $u^a$ in the following way
\begin{equation}
\label{8}
\delta u^a=\dot{\epsilon}^a-V_b^a\epsilon^b-u^c\epsilon^bC_{cb}^a+\partial_c\,(U_b^{ca} \epsilon^b)\,.
\end{equation}
Transformations of eqns. (\ref{7}) and (\ref{8}) constitute the gauge transformations
of the extended action of eqn. (\ref{5}) for any gauge functions 
$\epsilon^a=\epsilon^a(x^\mu)$ vanishing on the boundaries.

Gauge transformations of the total and original actions are determined by a partial 
gauge fixing of the extended action of eqn (\ref{5}). To demonstrate this, we label the constraints in a slightly different way 
so that it is manifest which generation each constraint belongs to.  If one chooses the gauge 
\begin{equation}
\label{9}
u^{m(i)}=0\,\,\,\,\,\,\,\,\,\,\,\,\,\,\,\,i\geq2\,,
\end{equation}
where $i$ refers to the generation of constraints, the extended action of eqn. (\ref{5}) reduces to the total action
 \begin{equation}
\label{10}
S_T=S_T\left(q^n,p_n,u^{m(1)}\right)=\int\! dt\, L_T
\end{equation} 
where 
\begin{equation}
\label{11}
L_T=\int d^mx\,\left(p_n{\dot q}^n-\mathcal{H}-u^{m(1)}\gamma_{m(1)}\right),
\end{equation} 
with only primary $1^{st}$class constraints present. Despite the gauge conditions 
of eqn. (\ref{9}), the total Lagrangian still possesses a residual gauge symmetry, 
which is the group of those symmetries of the extended action that preserve the gauge
condition of eqn. (\ref{9}); that is, gauge transformations for which

\begin{equation}
\label{12}
\delta u^{m(i)}=0\,\,\,\,\,\,\,\,\,\,\,\, i\geq2\,.
\end{equation}

Through eqn. (\ref{8}), this latter condition restricts the gauge functions $\epsilon^a=\epsilon^a(x^\mu)$ from being arbitrary 
and implies that they satisfy the differential equations,
\begin{equation}
\label{13}
\dot{\epsilon}^{\,m_i}-\sum_{j\geq i-1} V_{m_j}^{m_i}\epsilon^{\,m_j}-
u^{m_1}\sum_{j\geq i}C_{{m_1}{m_j}}^{m_i}\,\epsilon^{\,m_j}+
\sum_{j\geq i-1}\partial_k\,(U_{m_j}^{k\,m_i} \epsilon^{\,m_j})\,=0\,\,\,\,\,\,\,\,\,\,i\geq2\,,
\end{equation}
with summation over $j$ and $k$\,; $j,k=1 \ldots m$ implicit.

Provided a set of gauge functions $\epsilon^a=\epsilon^a(x^\mu)$ satisfy this set of equations,
 the residual transformations of eqn. (\ref{7})
constitute the gauge transformations of the total Lagrangian of eqn. (\ref{11}), and consequently of the original Lagrangian.    

\section{Application to Linearized Gravity}
Relabeling the constraints of eqns.\ (14), (15), (17) and (18) in the following way
\begin{equation}
\label{14}
\Phi_{0(1)}=\Pi\,\,\,\,\,\,\,\,\,\,\,\,\,\,\,\,\,\Phi_{0(2)}=\chi\,\,\,\,\,\,\,\,\,\,\,\,
\,\,\,\,\,\Phi_{i(1)}=\Pi_i\,\,\,\,\,\,\,\,\,\,\,\,\,\,\,\,\Phi_{i(2)}=\chi_i\,\,\,\,\,\,\,\,\,\,\,\,\,\,\,\,\,\,i=1\ldots n\,;
\end{equation}
(with numbers in parenthesis labeling the generation), the only nonzero constraint 
structure functions of eqns.\ (24) and (25) for linearized gravity are 
\begin{equation}
\label{15}
V_{0(1)}^{0(2)}=-1\,\,\,\,\,\,\,\,\,\,\,\,\,\,\,\,V_{i(1)}^{i(2)}=-1
\,\,\,\,\,\,\,\,\,\,\,\,\,\,\,\,\,U_{0(2)}^{k\,i(2)}=\frac{1}{2}\,\delta_{ki}
\,\,\,\,\,\,\,\,\,\,\,\,\,\,\,\,\,\,\,\,\,\,\,\,\,i,k=1 \ldots n\,.
\end{equation}
These, according to eqn. (\ref{13}), lead to the following equations for gauge functions $\epsilon^a$\,,  
\begin{eqnarray}
\label{16}
\dot \epsilon^{\,0(2)}+\epsilon^{\,0(1)} &=& 0\\
\dot \epsilon^{\,i(2)}+\epsilon^{\,i(1)}+\,\frac{1}{2}\,\partial_i \,\epsilon^{\,0(2)} &=&0\,\,.
\end{eqnarray}
We note that the upper indices $0$ and $i$ are merely {\it indices} for labeling gauge parameters, 
while the lower index $i$ stands for spatial coordinates. The gauge generator $G=\epsilon^a \gamma_a$ then is 
\begin{equation}
\label{17}
G=-\,\dot \epsilon^{\,0(2)}\,\Pi+\epsilon^{\,0(2)}\,\chi-
(\,\dot \epsilon^{\,i(2)}+\frac{1}{2}\,\partial_i\,\epsilon^{\,0(2)}\,)\,\Pi_i+\epsilon^{\,i(2)}\,\chi_i
\end{equation}
leading by eqn. (\ref{7}) to the following gauge transformations for the fields $\phi_{\mu\nu}$
\begin{equation}
\label{18}
\delta \phi=-\,\dot \epsilon^{\,0(2)}\,\,\,\,\,\,\,\,\,\,\,\,\,\,\delta 
\phi_i=-\,\dot \epsilon^{\,i(2)}-\frac{1}{2}\,\partial_i \,\epsilon^{\,0(2)}
\,\,\,\,\,\,\,\,\,\,\,\,\,\,\,\,\delta \phi_{ij}=-\,\partial_i \,\epsilon^{\,j(2)}
-\partial_j\,\epsilon^{\,i(2)}\,.
\end{equation}

If one introduces the covariant vector $\xi$ by  
$(\,\xi_0\,,\xi_i\,) \equiv (\,2\,\epsilon_{0(2)}\,,-\,\epsilon _{i(2)}\,)$\,, $i=1\ldots n$, one may then 
treat the indices of gauge functions as tempo-spatial indices and cast the gauge 
transformations of eqn. (\ref{18}) into the covariant form
\begin{equation}
\label{19}
\delta \phi_{\mu\nu}=\partial_\mu\,\xi_\nu+\partial_\nu\,\xi_\mu\,.
\end{equation}
This is the linearized diffeomorphism invariance of the original action of eqn. (\ref{006}) for a spin-two field.
\section{First Order Forms}
In analogy with the first order formaulation of electrodynamics, a first order formulation of the 
weak field action corresponding to (\ref{006}) by means of an auxiliary antisymmetric tensor field 
is possible. By adding an appropriate surface term, the Lagrangian density of eqn. (\ref{006}) can 
be cast into the following form
\begin{equation}
\label{4001}
\mathcal{L}=\frac{1}{4}\,(\partial_\mu \phi^\kappa_{\,\,\,\kappa}-\partial_\nu \phi^{\,\nu}_{\,\,\mu})\,
(\partial^\mu \phi^\kappa_{\,\,\,\kappa}-\partial_\kappa \phi^{\,\kappa \mu})-
\frac{1}{8}\,(\partial_\mu \phi_{\nu \kappa}-\partial_\nu \phi_{\mu \kappa})\,
(\partial^\mu \phi^{\nu \kappa}-\partial^\nu \phi^{\mu \kappa})\,.
\end{equation}
This suggests the introduction of the antisymmetric field
\begin{equation}
\label{4002}
G_{\mu \nu \kappa}=\partial_\mu \phi_{\nu \kappa}-\partial_\nu \phi_{\mu \kappa}\,\,;\,\,\,\,\,
\,\,\,\,\,\,\,\,\,\,\,\,\,\,\,\,\,\,\,\,\,\,\,\,\,\,\, G_{\mu \nu \kappa}=-G_{\nu \mu \kappa}
\end{equation} 
in terms of which the Lagrangian density of eqn. (\ref{006}) is
\begin{equation}
\label{4003}
\mathcal{L}=\frac{1}{4}\,G_{\mu\nu}^{\,\,\,\,\,\,\,\nu}\,G^{\,\mu\,\,\,\kappa}_{\,\,\,\kappa}-\frac{1}{8}\,
G_{\mu\nu\kappa}\,G^{\mu\nu\kappa}\,.
\end{equation}   
This in turn suggests the first order Lagrangian density
\begin{equation}
\label{4004}
\mathcal{L}=\frac{1}{8}\,G_{\mu\nu\kappa}\,G^{\mu\nu\kappa}-\frac{1}{4}\,G_{\mu\nu}^{\,\,\,\,\,\,\,\nu}\,
G^{\,\mu\,\,\,\kappa}_{\,\,\,\kappa}-\frac{1}{4}\,G^{\mu\nu\kappa}(\partial_\mu \phi_{\nu \kappa}-\partial_\nu \phi_{\mu \kappa})
+\frac{1}{2}\,G_{\mu\nu}^{\,\,\,\,\,\,\,\nu}(\partial^\mu \phi^\kappa_{\,\,\,\kappa}-\partial_\kappa \phi^{\mu\kappa})
\end{equation}
whose equivalence with the original Lagrangian can be verified by working out the equations of motion, and 
solving for the auxiliary field $G_{\mu\nu\kappa}$ in terms of the field $\phi_{\mu\nu}$\,.

However, a much more interesting first order formulation for the weak field action corresponding to (\ref{006}) is possible. It is related 
to the full Einstein-Hilbert action of eqns. (\ref{001}) and (\ref{002}) in first order form; that is, when 
$\Gamma^\lambda_{\mu\nu}$ and $g_{\mu\nu}$ are treated as independent fields. In the weak field limit 
$g_{\mu\nu}=\eta_{\mu\nu}+\phi_{\mu\nu}$ we have
\begin{eqnarray}
\label{4005}
\sqrt{-g}\,g^{\mu\nu} &=& (1+\frac{1}{2} \eta^{\alpha\beta}\phi_{\alpha\beta}+\ldots)\,(\eta^{\mu\nu}-\phi^{\mu\nu}+\ldots)\nonumber\\
 &=& \eta^{\mu\nu}-\phi^{\mu\nu}+\frac{1}{2}\, \eta^{\mu\nu}\,\phi^\sigma_{\,\,\sigma}+O(\phi^2)\,,
\end{eqnarray}
If we assume that in the first order weak field 
limit of the Einstein-Hilbert action the three index field $\Gamma^\lambda_{\mu\nu}$ appears with no change, and we also use 
the substitution of eqn. (\ref{4005}), then by keeping all terms bilinear in $\phi_{\mu\nu}$ and $\Gamma^\lambda_{\mu\nu}$, 
the Einstein-Hilbert 
Lagrangian density of eqn. (\ref{002}) becomes
\begin{equation}
\label{4006}
\mathcal{L}
=(\phi^{\mu\nu}-\frac{1}{2}\, \eta^{\mu\nu}\,\phi^\sigma_{\,\,\sigma})_{,\,\lambda}\,
\Gamma^\lambda_{\mu\nu}-\,(\phi^{\mu\nu}-\frac{1}{2}\, \eta^{\mu\nu}\,\phi^\sigma_{\,\,\sigma})_{,\,\nu}\,\Gamma^\lambda_{\lambda\mu}+
\eta^{\mu\nu}\left[\,\Gamma^\lambda_{\mu\nu}\Gamma^\sigma_{\sigma\lambda}-\Gamma^\lambda_{\sigma\mu}\Gamma^\sigma_{\lambda\nu}\,\right]\,.
\end{equation}  
(We have moved the derivatives on the three index field $\Gamma^\lambda_{\mu\nu}$ onto the perturbation 
field $\phi_{\mu\nu}$ by the addition of a surface term.) 

The action of eqn. (\ref{4006}) when $\Gamma^\lambda_{\mu\nu}$ and 
$\phi_{\mu\nu}$ are treated as independent fields is indeed the first order form of the action corresponding to (\ref{006})\,. 
This equivalence can be verified by working out the Lagrangian equations of motion for the three index field 
$\Gamma^\lambda_{\mu\nu}$ and solving for them in terms of the field $\phi_{\mu\nu}$. This 
first order formulation has the advantage of being also the weak field limit of the first order Eintein-Palatini 
formulation of the full Einstein-Hilbert action. It is hoped that a canonical analysis of the first order form of the spin-two field 
action will assist in giving an unambigous canonical treatment of the $(n+1)$ dimensional Einstein-Hilbert action in 
first order form, as it is expected that the structure of constraints be somewhat similar in both cases. 
We note that by ref. \cite{Gerry2} where
this procedure is followed in the two-dimensional action in first order form, and by ref. \cite{Ramin} where the 
two-dimensional metric and tetrad gravities are treated as constrained second order systems, it is possible that 
a gauge transformation which is distinct from diffeomorphism invariance may arise. This is currently being investigated.  

\vspace{2cm}
\noindent
{\bf Acknowledgments}
\vspace{.2cm}

\noindent The author wishes to thank Gerry McKeon for helpful discussions, suggestions and reading the 
manuscript. Kayvan Kargar is acknowledged for his encouragement.



\vspace{.1cm}


\begin{thebibliography}{99}

\bibitem{cliff} C. Burgess and G. Moore, {\it The Standard Model: A Primer} (Cambridge University Press, Cambridge 2006).

\bibitem{Gerry} D. G. C. McKeon, {\it Can. J. Phys.} {\bf 57} (1979) 2096


\bibitem{Henneaux} M. Henneaux, C. Teitleboim and J. Zanelli, {\it Nucl. Phys. B} {\bf 332} (1990) 169.

\bibitem{Henneaux2} M. Henneaux and C. Teitelboim, {\it Quantization of Gauge Systems}\, (Princeton U. Press, Princeton 1992).
 
\bibitem{Gerry2} N. Kiriushcheva, S. V.  Kuzmin and D.G. C. McKeon, {\it Int. J. Mod. Phys. A} {\bf 21} (2006) 3401-3420

\bibitem{Ramin} R. N. Ghalati, N. Kiriushcheva and S. V.  Kuzmin, {\it Mod. Phys. Lett. A} {\bf 22} {2007} 17-28

\end{thebibliography}
\end{document}